\documentclass[12pt]{article}
\usepackage[]{epsfig}
\usepackage[centertags]{amsmath}
\usepackage{amsfonts}
\usepackage{amssymb}
\usepackage[english]{babel}
\usepackage{graphicx}

\begin{document}

\title{\bf A note on the existence of soliton solutions in the  Chern-Simons-CP(1) model }
\author{
Lucas Sourrouille\thanks{\mbox{Corresponding author.}
\mbox{\emph{E-mail~address}:~sourrou@df.uba.ar (L. Sourrouille)}},
Alvaro Caso and Gustavo S Lozano,
\\
{\normalsize \it Departamento de F\'\i sica, FCEyN, Universidad
de Buenos Aires}
\\
{\normalsize \it Instituto de F\'\i sica de Buenos Aires, CONICET}
\\ {\normalsize\it Pab.1, Ciudad Universitaria,
1428, Ciudad de Buenos Aires, Argentina}
}

\maketitle

\abstract{We study a gauged Chern-Simons-CP(1) system.  We show that contrary to previous claims  the model in the absences of a potential term cannot support finite size soliton solution in $R^2$. }


\newpage


\vspace{1cm}

 The $CP(n)$ sigma model have been investigated in detail since the early 70's mainly as toy models to explore  the strong coupling effects of $QCD$ and as effective models of some condensed matter systems. In the $2+1$ dimensional case they have been argued to probably play a role in the description of High $T_c$ superconductors \cite{gral}.

An important issue related to this type of models concern the existence of soliton type solutions. For the simplest $CP(1)$ mode topological solutions have been shown to exist\cite{polyakov}.  Nevertheless, the solutions are of  arbitrary size due to scale invariance. As argued originally by Dzyaloshinsky, Polyakov and Wiegmann\cite{polyakov1} a Chern-Simons term can naturally arise in this type of models and the presence of a dimensional parameter could  play some role stabilizing the soliton solutions. A first detailed consideration of this problem was done in Ref.\cite{voru} where a perturbative analysis around the scale invariant solutions (i.e no Chern Simons coupling $\kappa=0$) showed that  the solutions were pushed to infinite size.
Still, it remained open the possibility of non perturbative solutions. This issue was considered in Ref.\cite{mehta} where by direct numerical analysis the authors claimed the existence of stable solitons. 

The aim of this note is to reconsider this problem in more detail and as a result of our investigation we will show  that contrary to the claim made in Ref.\cite{mehta}  that this solutions {\em do not} exist
in $R^2$. Still, we will show that the problem can be considered in a finite two dimensional disk, which could be a possible scenario in condensed matter applications (see \cite{gl} for the analogous problem of vortices in Ginzburg Landau theories)

We begin by considering a $(2+1)$ dimensional Chern-Simons model coupled to a complex two component field $n(x)$  described by the action

\begin{eqnarray}
S&=& S_{cs}+\int_{D} d^3 x |D_\mu n|^2
\label{S1}
\end{eqnarray}
The subindex $D$ indicates that the region of integration is a disc $D$ of radius $R$.
Here  $D_{\mu}= \partial_{\mu} + iA_{\mu}$ $(\mu =0,1,2)$ is the covariant derivative and $S_{cs}$ is the Chern-Simons action given by

\begin{eqnarray}
 S_{cs}= \kappa\int_{D} d^3 x \epsilon^{\mu \nu \rho} A_\mu \partial_\nu A_\rho
\end{eqnarray}
where
\begin{eqnarray}
F_{\mu \nu}=\partial_{\mu}A_{\nu}-
\partial_{\nu}A_{\mu} \label{F}
\end{eqnarray}%
The signature of the metric $g^{\mu \nu}$ is $(1,-1,-1)$ and  the two component field $n(x)$ is subject to
the constraint $n^\dagger n = 1$

Variation of the action yields the field equations

\begin{eqnarray}
D_\mu D^\mu n = (n^\dagger D_\mu D^\mu n)n
\label{motion1}
\end{eqnarray}

\begin{equation}
 \kappa\epsilon_{\mu \nu \rho} F^{ \nu \rho} = - J_\mu=i [n^\dagger D_\mu n - n(D_\mu n)^\dagger]
\end{equation}
The time component of last equation
\begin{eqnarray}
2\kappa F_{12} =  -J_0 \label{gauss}
\end{eqnarray}
is  Gauss's law of Chern-Simons dynamics. Integration  over the disc gives a relation  between  charge $Q =\int_{D} d^2 x J_0$ and magnetic flux $\Phi = \int_{D} F_{12} d^2 x$ \cite{Echarge}:
\begin{eqnarray}
\Phi = -\frac{1}{2\kappa} Q
\label{Q}
\end{eqnarray}
Defining the stress tensor as $T_{ \mu \nu}=\frac{\delta S}{\delta g^{\mu \nu}}$ and using the equations of motion, the energy functional for a static field configuration can be expressed as
\begin{eqnarray}
E=  \int_{D} d^2 x \Big(\kappa^2 B^2 + |D_i n|^2  \Big) \,,
\;\;\;\;\;\
i = 1,2  \label{statich}
\end{eqnarray}

Let us consider the following ansatz for the $N$ soliton solutions \cite{mehta}:

\begin{eqnarray}
n(\phi, r)=  \left( \begin{array}{c}
\cos(\frac{\theta(r)}{2})e^{i N \phi}\\
\sin(\frac{\theta(r)}{2} )\end{array} \right)
\,,
\;\;\;\;\;\
 A_\phi (r)= a(r)
\,,
\;\;\;\;\;\
A_r =0
\label{ansatz}
\end{eqnarray}

Using this ansatz Eq.(\ref{statich}) becomes

\begin{eqnarray}
E= 2\pi \int_0^R r dr \Big(\kappa^2 \left( \frac{a(r)}{r}+ \partial_r a(r) \right)^2 +\frac{1}{4}(\partial_r \theta(r))^2 \nonumber \\
+ \left(\frac{N^2}{r^2} + \frac{2Na(r)}{r}\right)\cos^2(\frac{\theta(r)}{2})+ a^2(r)  \Big) \,
\;\;\;\;\;\
\label{statich1}
\end{eqnarray}

We shall impose the following boundary conditions at the origin
\begin{eqnarray}
\lim_{r \to 0} \theta(r) = \pi
\,,
\;\;\;\;\;\
\lim_{r \to 0} a(r) = 0
\label{b1}
\end{eqnarray}
These conditions imply regularity of the fields at the origin.
On the other hand the conditions at the boundary of the disk are in principle more general. If we were working on the infinite plane, then the natural boundary conditions ensuring finite energy would be,

\begin{eqnarray}
\lim_{r \to \infty} \theta(r) = 0
\,,
\;\;\;\;\;\
\lim_{r \to \infty}a(r) =-\frac{ N}{r}
\label{b2}
\end{eqnarray}
For a finite disk it seems reasonable to impose
\begin{equation} 
\lim_{r \to R}a(r) =-\frac{ N}{R}
\end{equation}
which implies quantization of the magnetic flux
\begin{eqnarray}
\Phi =  2\pi\int_0^R r dr\,\, \frac{\partial_r(r\,\,a(r))}{r}=-2\pi N
\label{flux}
\end{eqnarray}
For the $\theta$ field we could impose a Dirichlet type boundary condition,
\begin{equation}
\lim_{r \to R}\theta(r)=\theta(R) 
\label{dirich}
\end{equation}
or a Neumann type boundary condition
\begin{equation}
\lim_{r \to R}\frac{d\theta(r)}{dr}=\theta^{'}(R) 
\label{neu}
\end{equation}
The condition $\theta(R)=0$ is special in the sense that it implies the vanishing of the current at the boundary,
\begin{equation}
J_\phi = \Big(-1 + \cos^2(\frac{\theta(R)}{2})\Big)\frac{2N}{R}
\label{current}
\end{equation}

Using the equations of motion, it can be shown that,
\begin{eqnarray}
\nabla . {\bf E} = \frac{1}{2\kappa^2}J_0 + \frac{2\pi}{\kappa} J^{CP(1)}_0
\label{Div}
\end{eqnarray}
where the $CP(1)$ charge density is defined as

\begin{equation}
J^{CP(1)}_0=\frac{i}{2\pi}\epsilon^{i j} (D_i n)^\dagger (D^j n)
\end{equation}
Then,
\begin{eqnarray}
B =  2\pi J^{CP(1)}_0 - \kappa\nabla . {\bf E} \label{B2}
\end{eqnarray}
Using that,
\begin{equation}
2\kappa\nabla . {\bf E} = -(\partial_1 J_2 -\partial_2 J_1)
\end{equation}
\begin{eqnarray}
\Phi = \int_{D} B \,\, d^2 x = 2 \pi \int_{D} J^{CP(1)}\,\, d^2 x +\frac{1}{2} \int_{\partial D} J_i dx^{i}
\label{M}
\end{eqnarray}
We clearly see that the relation between magnetic flux and $CP(1)$ charge involves a boundary term related to the current.

Using the Bogomol'nyi identity \cite{bogo} $|D_i n|^2 =   |( D_1 \pm iD_2)n|^2 \mp B \pm \frac{\epsilon^{ij}}{2} \partial_i J_j$ , the energy (\ref{statich}) becomes

\begin{eqnarray}
E=\int_{DR} d^2 x \,\, \Big( \kappa^2 B^2 +  |( D_1 \pm iD_2)n|^2  \Big) \mp 2\pi Q_{CP(1)}
\label{H2}
\end{eqnarray}
We see that the energy is bounded bellow by the $CP(1)$ charge.
\begin{eqnarray}
E \geq  2\pi \int_{DR} d^2 x \,\, J_{CP(1)}^0 =  2\pi  |Q_{CP(1)}|
\label{H}
\end{eqnarray}

This bound is saturated by fields satisfying the first-order Bogomol'nyi self-duality equations\cite{bogo}.

\begin{eqnarray}
|( D_1 \pm iD_2)n|^2 =0
\,,
\;\;\;\;\;\
B=0
\label{bogo}
\end{eqnarray}

The field equations corresponding to the ansatz Eq.(\ref{ansatz}) read

\begin{eqnarray}
\partial_r^2 a(r)+\frac{\partial_r a(r)}{r}- \frac{a(r)}{r^2} - \frac{a(r)}{\kappa^2}=\cos^2(\frac{\theta(r)}{2})\frac{N}{r\kappa^2}
\label{m1}
\end{eqnarray}
\begin{eqnarray}
r\partial_r(r\partial_r \theta(r)) + N^2 \sin(\theta(r)) = -2Nra(r)\sin(\theta(r))
\label{m2}
\end{eqnarray}
If the solutions of (\ref{m1}) and (\ref{m2}) exist their scale must be set by the quantity $\kappa$. Following  Ref.\cite{mehta}, we introduce the dimensionless quantities

\begin{eqnarray}
A = \kappa a
\,,
\;\;\;\;\;\
s =\frac{r}{\kappa}
\end{eqnarray}
in terms of which (\ref{m1}) and (\ref{m2}) become

\begin{eqnarray}
\partial_s^2 A +\frac{\partial_s A}{s}- \frac{A}{s^2} - A =\cos^2(\frac{\theta}{2})\frac{N}{s}
\label{m11}
\end{eqnarray}
\begin{eqnarray}
s\partial_s(s\partial_s \theta ) + N^2 \sin(\theta ) = -2NsA\sin(\theta)
\label{m22}
\end{eqnarray}

The energy functional (\ref{statich1})  in terms of these new variables reads as

\begin{eqnarray}
E(S)= 2\pi \int_0^S s ds \Big( \left( \frac{A}{s}+ \partial_s A \right)^2 +\frac{1}{4}(\partial_s \theta)^2 \nonumber \\
+ \left(\frac{N^2}{s^2} + \frac{2NA}{s}\right)\cos^2(\frac{\theta}{2})+ A^2  \Big) \,
\;\;\;\;\;\
\label{H3}
\end{eqnarray}
For the origin we choose the following boundary conditions,
\begin{eqnarray}
\lim_{s \to 0} \theta = \pi
\,,
\;\;\;\;\;\
\lim_{s \to 0} A = 0
\label{b11}
\end{eqnarray}
while for the boundary $S=R/\kappa$ we choose,

\begin{eqnarray}
& &\lim_{s \to S}A =-\frac{ N}{S} \\
& & \lim_{s \to S} \theta =\theta_o  \,\,\,\, or \,\,\,\,\  \lim_{s \to S} \theta^{'}  =\theta^{'} (R)
\,,
\;\;\;\;\;\
\label{b22}
\end{eqnarray}

The field equations (\ref{m11}) and (\ref{m22}) are the equations presented in $(17)$ and $(18)$  of Ref.\cite{mehta}. The solutions to these equations can be analyzed numerically. From the numerical point of view, even if one were interested in the infinite plane, the equations have to be solved in a finite disk and eventually analyse the behavior for $R \to \infty$ .

There are basically two methods to solve numerically a boundary value problem  for a system of ordinary differential equation, generally referred as the ``shooting''and ``relaxation'' methods (see Ref.\cite{reci}). The first one is easier to implement but has more limitations. It consists essentially in guessing the conditions at one boundary to reproduce, after a standard numerical integration (like Euler or Runge-Kutta),  the expected behavior at the other boundary. This method is nevertheless of limited usefulness in problems with exponentially divergent solutions (as the problem at hand). This is so because an infinitesimal error at the initial condition exponentiates to the wrong behavior at the other end. Thus, there exists a maximum radius R  (which depends on machine precision) that can be explored using this approach. The authors of Ref. \cite{mehta} have used a variation of this method to study these equations. 

Indeed, the method implemented in Ref. \cite{mehta}, works as follows. One first takes 
\begin{equation}
\theta^{(0)}(s)=2 \arctan(\frac{1}{s})
\end{equation}
to replace at rhs of Eq (\ref{m11}), and solves for $A(s)$, choosing the value $A^{'(0)}(0)$ in order to reproduce the the expected behavior at $s=S$. With this solution, called $A^{(0)}(s)$ one goes to Eq (\ref{m22}) to solve for $\theta(s)$ giving a solutions that we call $\theta^{(1)}(s)$. This iterations is carried over on many steps, obtaining solutions, $A^{(i)}(s), \theta^{(i)}(s)$ in each case one  need to adjust the value of $A^{'(i)}(0),\theta^{'(i)}(0)$.
The advantage of this method over a standard ``shooting'' approach is that here one needs to guess one initial condition per iteration instead of trying to guess two boundary conditions. As mentioned before, due to the presence of exponentially divergent solutions at infinity, ``very fine tuning'' in $A^{'(i)}(0)$ is needed to reproduce the expected $-N/s$ behavior at ``infinity''.
In their paper, authors quote results for $N=1,S=30$.

\begin{figure}
\centering   
\includegraphics
[height=70mm,clip]{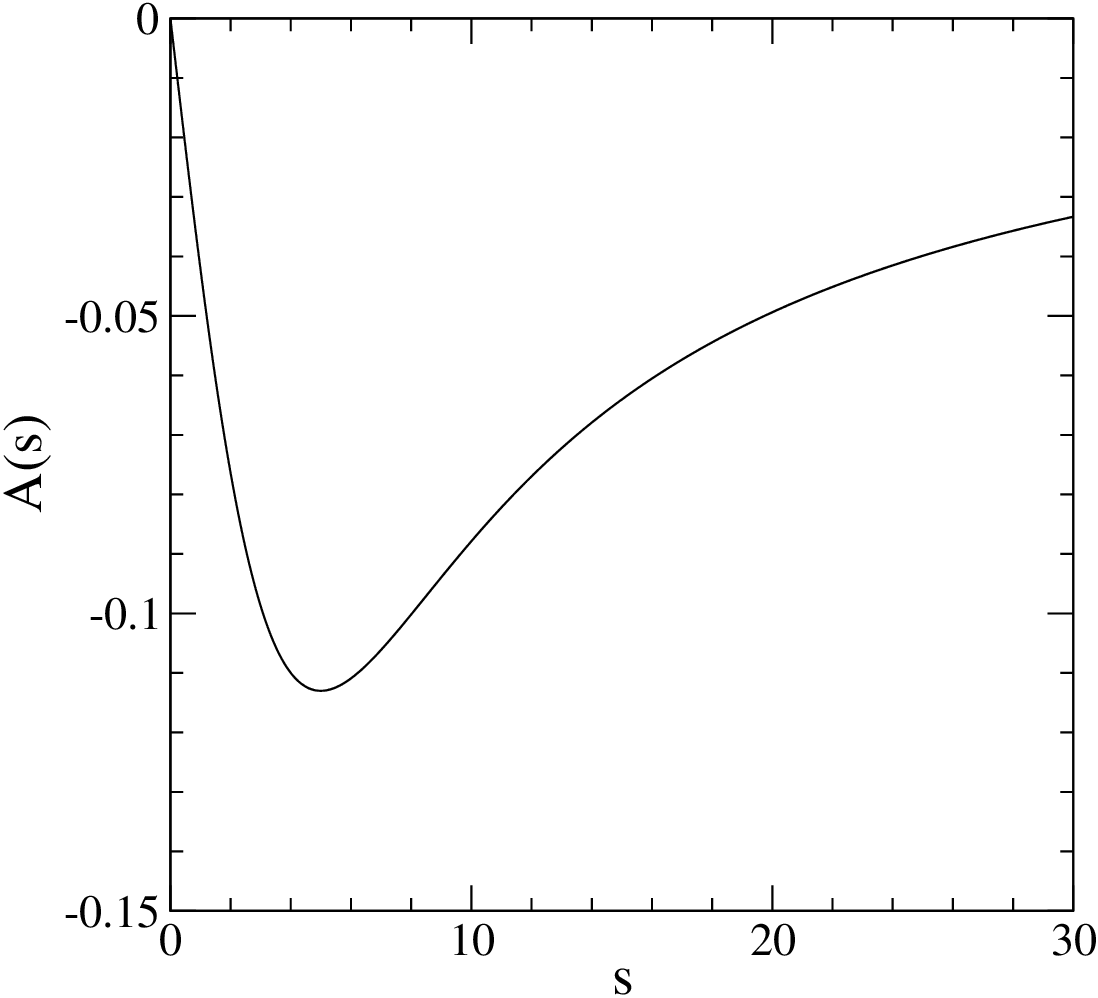}
\includegraphics
[height=70mm,clip]{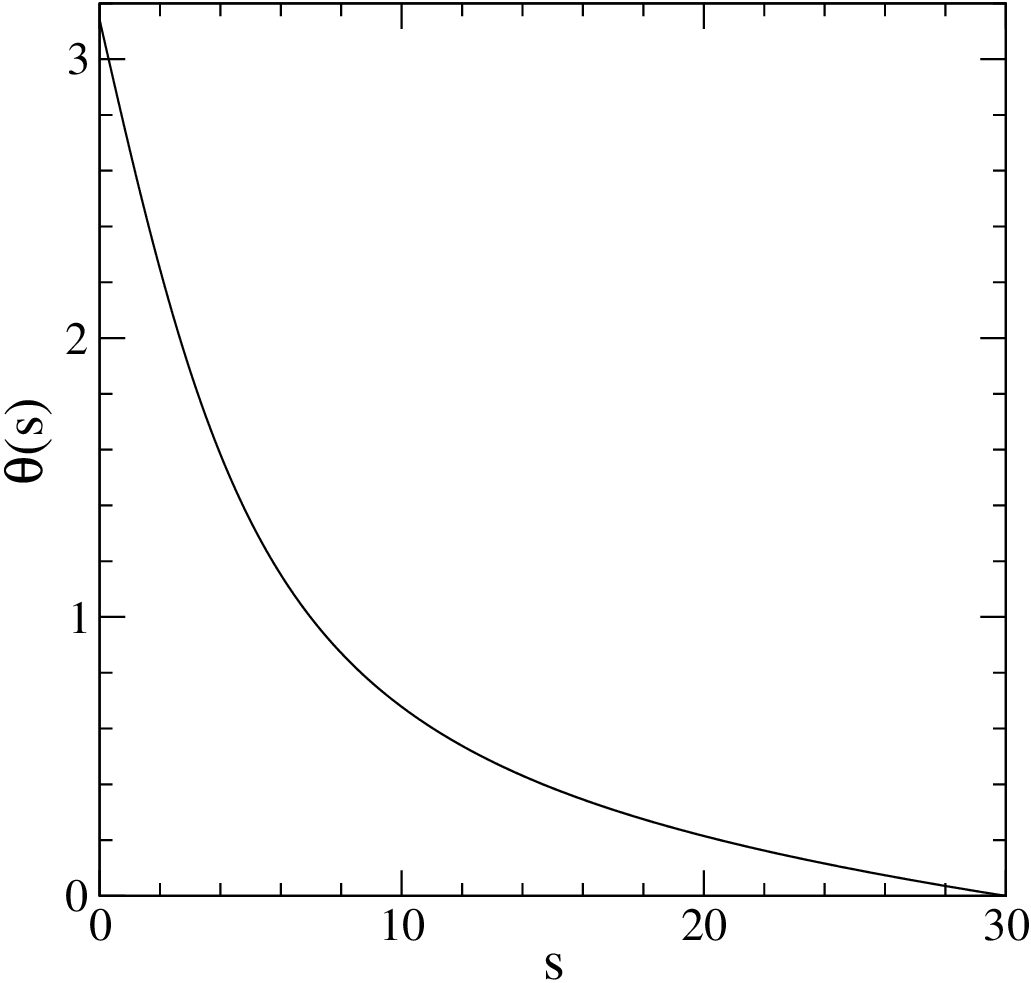} 
\caption{{
Profile of the gauge field $A(s)$ (left panel) and $\theta(s)$ (right panel) for disk size $S=30$ and $N=1$. 
}}
\label{tt}
\end{figure}

The same boundary value problem can be analyzed more efficiently using a ``relaxation'' method (see for instance Ref.\cite{reci} for details about its implementation).
In this case, arbitrary disk sizes can be explored since the routine works trying to achieve global convergence starting from an initial guess for the solutions in the full interval. 

We have solved the problem using both methods, the one proposed in 
Ref. \cite{mehta} and the relaxation method mentioned before, for $N=1$ and $S=30$, imposing as boundary condition $a(S)=-\frac{1}{S},\theta(S)=0$. 

In Fig \ref{tt}  we show the solutions for the gauge field obtained by both methods. After 30 iterations of the relaxation method, the agreement between the solutions obtained by both methods is excellent and are indistinguishable in the graph. They also look compatible with   the results presented in Ref. \cite{mehta}.

\begin{figure}
\centering   
\includegraphics
[height=70mm]{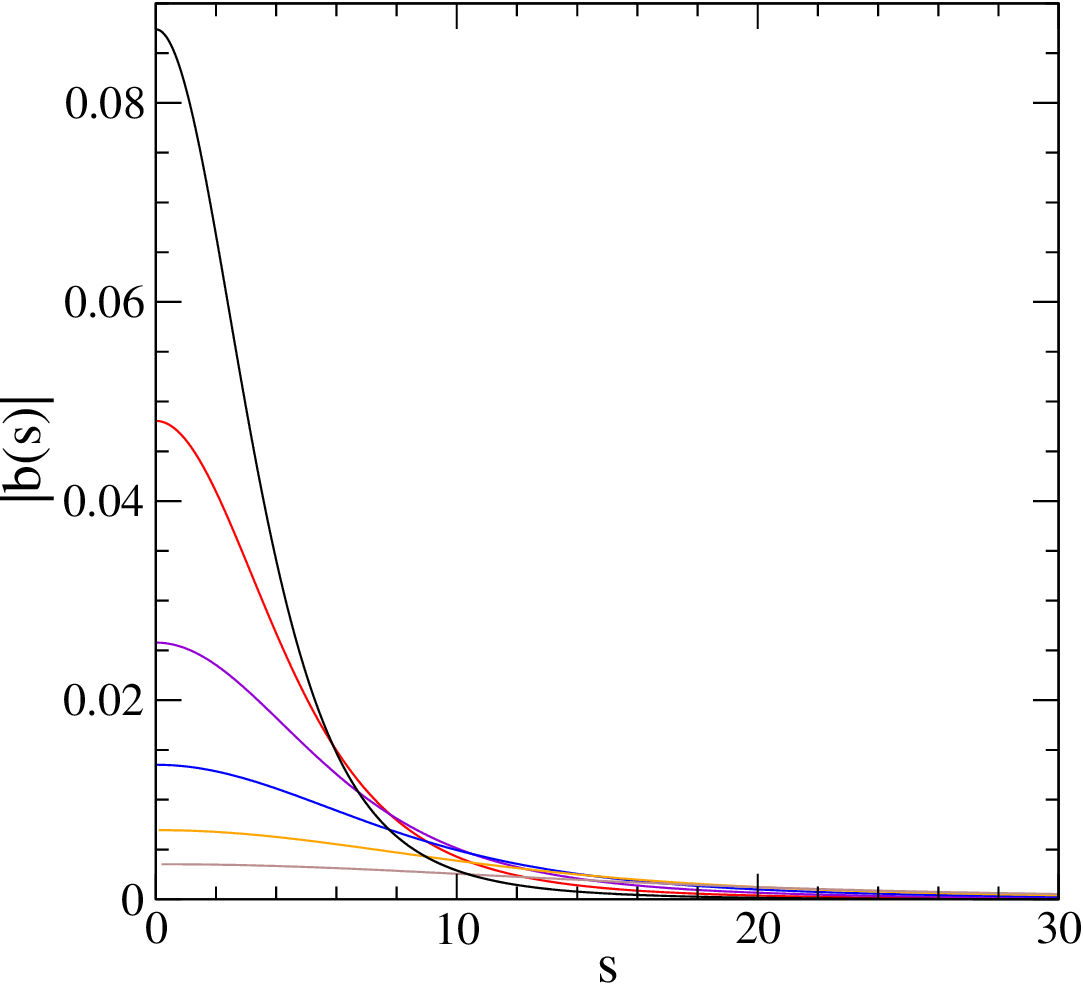} 
\caption{{
(Color online)`
{Magnetic field profile as a function of scaled radial coordinate $s$ for different disk sizes, from top to bottom, $S=30,60,120,240,480,960$}
}}
\label{tt2}
\end{figure}

As mentioned before, the shooting method becomes inefficient for larger radius while the relaxation method we can be use to analyse the behavior of the solutions as a function of disk size  $S$.
In Fig \ref{tt2} we show the behavior of the solutions for different disk sizes.
This figure clearly shows that as the disk sizes increases the solution tends to the trivial solutions. Our results clearly contradict the conclusion of Ref. \cite{mehta} that claim the existence of non trivial stable finite energy  Chern-Simons CP(1) solitons in $R^2$. Indeed, our result points in the direction that conclusion  of Ref. \cite{voru}, where instability was shown at the perturbative level can be extended to the non perturbative case too.

\begin{figure}
\centering   
\includegraphics
[height=70mm]{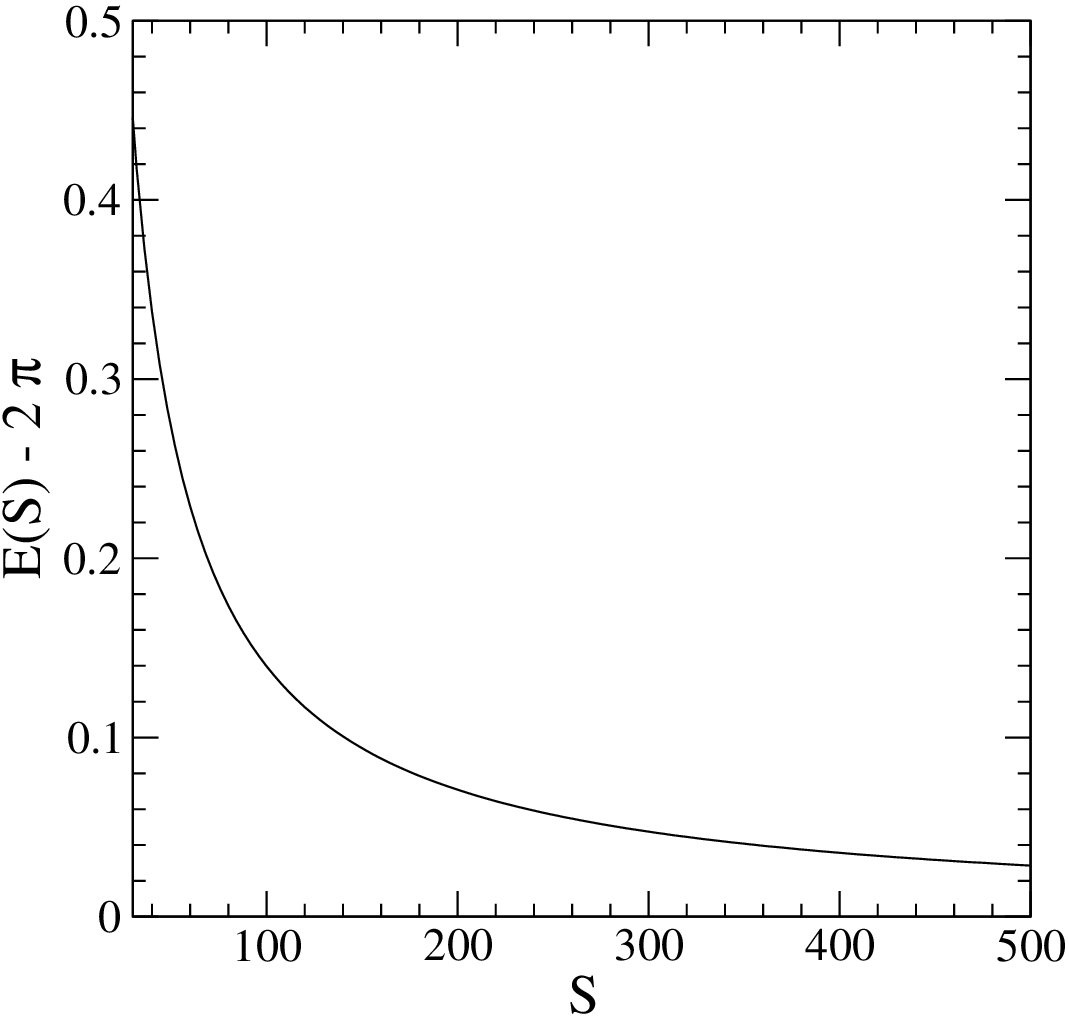}
\caption{{
The energy of the solution as a function of  disk size $S$, for $N=1$
}}
\label{tt3}
\end{figure}

The fact that the solutions delocalise as $S$ is increased could be related to the fact that the energy 
decreases with size. In Fig.\ref{tt3} we show the behavior of the energy as a function of the disk size, for $N=1$. This result can be shown also analytically. Indeed, 
consider the following configuration defined in the interval of length $\lambda S$

\begin{eqnarray}
\tilde{A}_{\lambda S}(s) =\frac{A_{ S}(\frac{s}{\lambda})}{\lambda}
\,,
\;\;\;\;\;\
\tilde{\theta}_{\lambda S}(s) = \theta_{S}(\frac{s}{\lambda})
\label{config}
\end{eqnarray}
Here $\lambda$ is a real number such that $\lambda >1$ and the configurations (\ref{config}) satisfy the boundary conditions
$
\left.\tilde{A}_{\lambda S}(s)\right|_{s=\lambda S} = -\frac{N}{\lambda S}
\,,
\;\;\;\;\;\
\left.\tilde{\theta}_{\lambda S}(s)\right|_{s=\lambda S} =0
$
We can evaluate the energy functional  (\ref{H3}) for the configuration (\ref{config}) in an interval of length $\lambda S$
\begin{eqnarray}
\tilde{E}(\lambda S)= 2\pi \int_0^{\lambda S} s ds \Big( \left( \frac{\tilde{A}_{\lambda S}(s)}{s}+ \partial_s \tilde{A}_{\lambda S}(s) \right)^2 +\frac{1}{4}(\partial_s \tilde{\theta})^2 \nonumber \\
+ \left(\frac{N^2}{s^2} + \frac{2N\tilde{A}_{\lambda S}(s)}{s}\right)\cos^2(\frac{\tilde{\theta}}{2})+ \tilde{A}^2_{\lambda S}(s)  \Big) \,
\;\;\;\;\;\
\label{H4}
\end{eqnarray}
We denote the solution corresponding to the interval $\lambda S$ as $A_{\lambda S} (s)$  and $\theta_{\lambda S} (s)$ and its energy as $E(\lambda S)$. Since the configuration (\ref{config}) satisfy the same boundary condition  as $A_{\lambda S} (s)$  and $\theta_{\lambda S} (s)$, we have that
\begin{eqnarray}
E(\lambda S) \leq \tilde{E}(\lambda S)
\label{inq}
\end{eqnarray}
Under the transformation $s= x\lambda$ the functional (\ref{H4}) becomes
\begin{eqnarray}
\tilde{E}(\lambda S)= 2\pi \int_0^{S} x dx \Big(\frac{1}{{\lambda}^2} \left( \frac{A_{S}(x)}{x}+ \partial_x A_{S}(x) \right)^2 +\frac{1}{4}(\partial_x \theta)^2 \nonumber \\
+ \left(\frac{N^2}{x^2} + \frac{2NA_{S}(x)}{x}\right)\cos^2(\frac{\theta}{2})+ A^2_{S}(x)  \Big) \,
\;\;\;\;\;\
\label{}
\end{eqnarray}

The last expression differ from $E(S)$ only on the factor $\frac{1}{\lambda^2}$, therefore
\begin{eqnarray}
\tilde{E}(\lambda S)< E(S)
\label{inq1}
\end{eqnarray}
Comparing (\ref{inq}) and  (\ref{inq1}) we have that
\begin{eqnarray}
E(\lambda S)< E(S)
\end{eqnarray}
that is,  the energy decreases when we enlarge the interval $S$. 

\begin{figure}
\centering   
\includegraphics
[height=70mm]{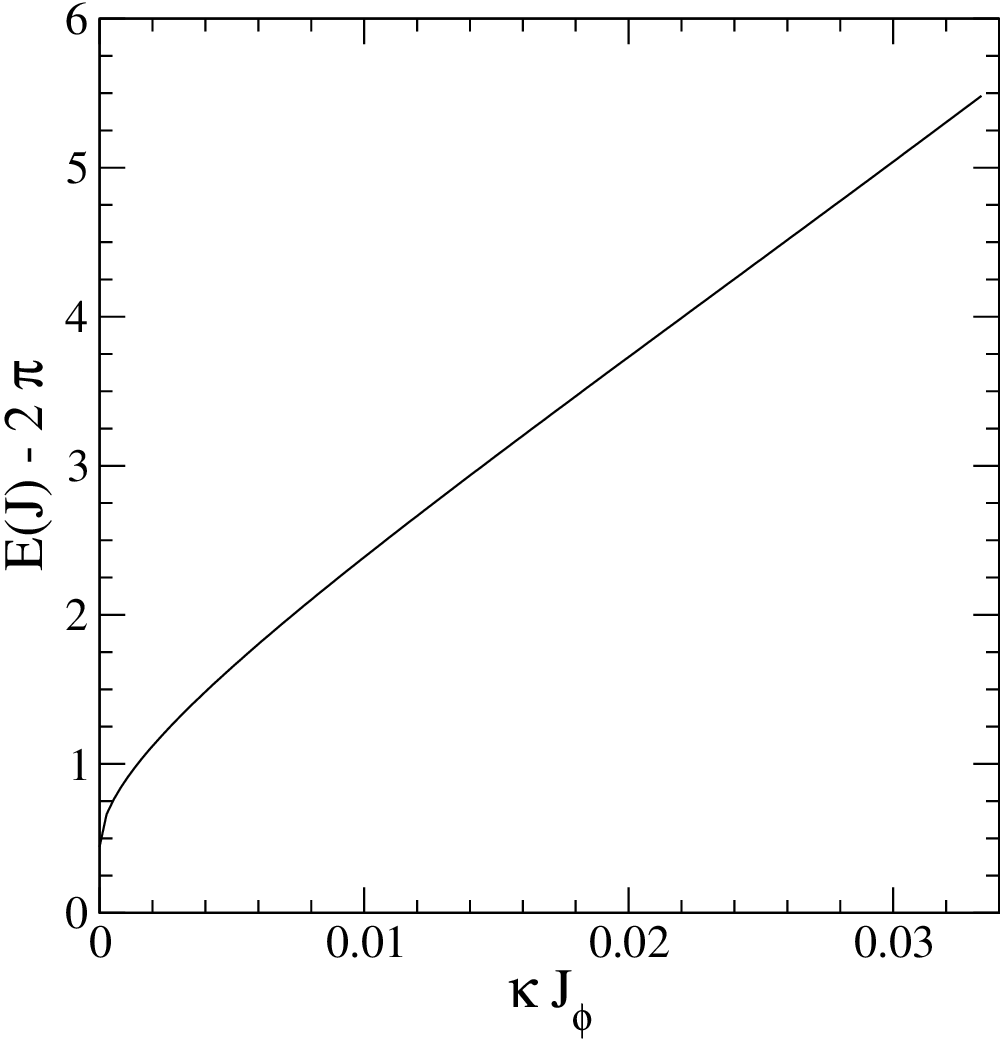} 
\caption{{
The energy as a function of the current at the boundary, for $S=30$ and $N=1$.
}}
\label{tt4}
\end{figure}

As we mentioned before, in a finite disk there is some freedom on the boundary conditions imposed on the fields. The conditions we have been using below, $\theta(S)=0$ is such the current $J_{\phi}(S)=0$ (see Eq. \ref{current}).
The behavior of the energy for more general conditions can be easily explored. We show in Fig \ref{tt4}, the energy as a function of the current at the boundary.

We would like to mention that the solutions are stabilized if additional potential term as in \cite{sigmas}, \cite{z} are included. In particular, we have checked the results of reference \cite{z} using the relaxation method and we have found results in complete agreement with them.

In this note then analyzed in more detail the properties of non trivial classical solutions in the $2+1$ CP(1)-Chern-Simons model. Contrary to previous claims, we have shown that these solutions are not stable in the infinite plane case, thus, extending to the non perturbative domain the conclusions of Ref. \cite{voru}.
The solutions are perfectly well defined in a finite disk and this could be the case of interest specially if thinking in condensed matter applications. The behavior of the solutions under different external conditions can be easily obtained following the methods described in this work.

{\bf Acknowledgements}

This work was partially supported through grant X123 of the University of Buenos Aires and PIP 1212/09 of Conicet, Argentina. We thank D. Perez Daroca for comments on the numercial analysis.

\end{document}